\documentclass[12pt]{article}
\usepackage{graphicx}
\usepackage{epsfig}
\usepackage{amsmath}
\usepackage{amsfonts}
\usepackage{amssymb}
\renewcommand{\baselinestretch}{2.0}

\def \pbarp {$p \bar p~$}
\def \roots {$\sqrt{s}$}

\def\zz {$\tilde{\chi}^2$}

\def\es {\epsilon_\gamma}
\def\eb {\epsilon_b}

\begin{document}
\begin{titlepage}
\begin{flushright}
FERMILAB-PUB-01/390-E\\
\end{flushright}
\vspace{2cm}
\begin{center}
{\LARGE Comparison of the Isolated
Direct Photon Cross Sections in $p \bar p$ Collisions at  $\sqrt{s}=1.8$ TeV and  $\sqrt{s}=0.63$ TeV}
\vspace{0.25cm}
{ \ }
{ \ }
\vspace{0.75cm}
\font\eightit=cmti8
\def\r#1{\ignorespaces $^{#1}$}
\hfilneg
\begin{sloppypar}
\noindent
D.~Acosta,\r {12} T.~Affolder,\r {23} H.~Akimoto,\r {45}
A.~Akopian,\r {37} M.~G.~Albrow,\r {11} P.~Amaral,\r 8  
D.~Amidei,\r {25} K.~Anikeev,\r {24} J.~Antos,\r 1 
G.~Apollinari,\r {11}
% ...
\end{sloppypar}
\vskip .026in
\begin{center}
(CDF Collaboration)
\end{center}

\vskip .026in
\begin{center}
\r 1  {\eightit Institute of Physics, Academia Sinica, Taipei, Taiwan 11529, 
Republic of China} \\
\r 2  {\eightit Argonne National Laboratory, Argonne, Illinois 60439} \\
\r 3  {\eightit Istituto Nazionale di Fisica Nucleare, University of Bologna,
I-40127 Bologna, Italy} \\
\r 4  {\eightit Brandeis University, Waltham, Massachusetts 02254} \\
\r 5  {\eightit University of California at Davis, Davis, California  95616} \\
\r 6  {\eightit University of California at Los Angeles, Los 
Angeles, California  90024} \\  
\r 7  {\eightit Instituto de Fisica de Cantabria, CSIC-University of Cantabria, 
39005 Santander, Spain} \\
\r 8  {\eightit Enrico Fermi Institute, University of Chicago, Chicago, 
Illinois 60637} \\
\r 9  {\eightit Joint Institute for Nuclear Research, RU-141980 Dubna, Russia}
\\
\r {10} {\eightit Duke University, Durham, North Carolina  27708} \\
\r {11} {\eightit Fermi National Accelerator Laboratory, Batavia, Illinois 
60510} \\
\r {12} {\eightit University of Florida, Gainesville, Florida  32611} \\
\r {13} {\eightit Laboratori Nazionali di Frascati, Istituto Nazionale di Fisica
               Nucleare, I-00044 Frascati, Italy} \\
\r {14} {\eightit University of Geneva, CH-1211 Geneva 4, Switzerland} \\
\r {15} {\eightit Glasgow University, Glasgow G12 8QQ, United Kingdom}\\
\r {16} {\eightit Harvard University, Cambridge, Massachusetts 02138} \\
\r {17} {\eightit Hiroshima University, Higashi-Hiroshima 724, Japan} \\
\r {18} {\eightit University of Illinois, Urbana, Illinois 61801} \\
\r {19} {\eightit The Johns Hopkins University, Baltimore, Maryland 21218} \\
\r {20} {\eightit Institut f\"{u}r Experimentelle Kernphysik, 
Universit\"{a}t Karlsruhe, 76128 Karlsruhe, Germany} \\
\r {21} {\eightit Center for High Energy Physics: Kyungpook National
University, Taegu 702-701; Seoul National University, Seoul 151-742; and
SungKyunKwan University, Suwon 440-746; Korea} \\
\r {22} {\eightit High Energy Accelerator Research Organization (KEK), Tsukuba, 
Ibaraki 305, Japan} \\
\r {23} {\eightit Ernest Orlando Lawrence Berkeley National Laboratory, 
Berkeley, California 94720} \\
\r {24} {\eightit Massachusetts Institute of Technology, Cambridge,
Massachusetts  02139} \\   
\r {25} {\eightit University of Michigan, Ann Arbor, Michigan 48109} \\
\r {26} {\eightit Michigan State University, East Lansing, Michigan  48824} \\
\r {27} {\eightit University of New Mexico, Albuquerque, New Mexico 87131} \\
\r {28} {\eightit The Ohio State University, Columbus, Ohio  43210} \\
\r {29} {\eightit Osaka City University, Osaka 588, Japan} \\
\r {30} {\eightit University of Oxford, Oxford OX1 3RH, United Kingdom} \\
\r {31} {\eightit Universita di Padova, Istituto Nazionale di Fisica 
          Nucleare, Sezione di Padova, I-35131 Padova, Italy} \\
\r {32} {\eightit University of Pennsylvania, Philadelphia, 
        Pennsylvania 19104} \\   
\r {33} {\eightit Istituto Nazionale di Fisica Nucleare, University and Scuola
               Normale Superiore of Pisa, I-56100 Pisa, Italy} \\
\r {34} {\eightit University of Pittsburgh, Pittsburgh, Pennsylvania 15260} \\
\r {35} {\eightit Purdue University, West Lafayette, Indiana 47907} \\
\r {36} {\eightit University of Rochester, Rochester, New York 14627} \\
\r {37} {\eightit Rockefeller University, New York, New York 10021} \\
\r {38} {\eightit Rutgers University, Piscataway, New Jersey 08855} \\
\r {39} {\eightit Texas A\&M University, College Station, Texas 77843} \\
\r {40} {\eightit Texas Tech University, Lubbock, Texas 79409} \\
\r {41} {\eightit Institute of Particle Physics, University of Toronto, Toronto
M5S 1A7, Canada} \\
\r {42} {\eightit Istituto Nazionale di Fisica Nucleare, University of Trieste/
Udine, Italy} \\
\r {43} {\eightit University of Tsukuba, Tsukuba, Ibaraki 305, Japan} \\
\r {44} {\eightit Tufts University, Medford, Massachusetts 02155} \\
\r {45} {\eightit Waseda University, Tokyo 169, Japan} \\
\r {46} {\eightit University of Wisconsin, Madison, Wisconsin 53706} \\
\r {47} {\eightit Yale University, New Haven, Connecticut 06520} \\
\r {(\ast)} {\eightit Now at Carnegie Mellon University, Pittsburgh,
Pennsylvania  15213} \\
\r {(\ast\ast)} {\eightit Now at Northwestern University, Evanston, Illinois 
60208} \\
\r {(\ast\ast\ast)} {\eightit Now at University of California, Santa Barbara, CA
93106}
\end{center}

\vspace{0.5cm}
\vfill
\begin{abstract}
We have measured the cross sections $d^2\sigma/dP_T d\eta$ for
production of isolated direct photons in \pbarp collisions at
two different center-of-mass energies, 1.8 TeV and 0.63 TeV, using the
Collider Detector at Fermilab (CDF).  The normalization of 
both data sets agree with the predictions of Quantum Chromodynamics (QCD) 
for photon transverse momentum ($P_T$) of 25 GeV/c,  but the shapes 
versus photon $P_T$ do not.  These shape differences lead 
to a significant disagreement in the ratio of cross sections
in the scaling variable $x_T (\equiv 2P_T/\sqrt{s}$).  This disagreement 
in the $x_T$ ratio is difficult to explain with conventional 
theoretical uncertainties such as scale dependence and 
parton distribution parameterizations.  

\end{abstract}
\end{center}
\vfill

\end{titlepage}
\clearpage
\newpage
\section{Introduction}
In this article we present a measurement of the cross section for
production of isolated prompt photons in
proton-antiproton collisions at center-of-mass energies 
\mbox{$\sqrt{s}=1.8$} TeV and 
\mbox{$\sqrt{s}=0.63$} TeV using the Collider Detector at Fermilab (CDF).
Prompt photons are those produced in the $p \bar{p}$ collision, in 
distinction from the copious background of photons produced by decays
of hadrons such as the $\pi ^0$ and $\eta$ mesons.  In
Quantum Chromodynamics (QCD), at lowest order, prompt photon
production is dominated by the Compton process $gq \rightarrow \gamma
q$, which is sensitive to the gluon distribution of the
proton~\cite{photon}.  This process has been previously measured by
the CDF collaboration 
at \roots = 1.8 TeV ~\cite{pho89,pho93}; a difference between the 
data and the theory was observed in the shape of the differential
cross section versus photon $P_T$.   The discrepancy
could possibly be eliminated by a change in the gluon distribution 
inside the proton~\cite{gluon_explanation}. In this analysis the photon cross
section has been measured using the same detector and the same
technique at two widely different center-of-mass energies, with a
larger data sample than that of the previous measurement.  
A comparison of the ratio of the two cross sections 
provides a precise test of the QCD calculations, as many of 
the experimental uncertainties cancel.
The cross section 
ratio also provides a direct probe of the
QCD matrix elements, as the theoretical 
uncertainties due to the gluon distribution are reduced.


%
%

\section{Detector Description and Event Selection}
A detailed description of the CDF detector may be found 
in Refs.~\cite{pho93,detcdf}. 
Here we describe the two 
detector systems used to distinguish prompt photons from 
neutral hadron backgrounds.
A multiwire proportional chamber with cathode
strip readout (the Central Electromagnetic Strip system, CES) embedded in
each central electromagnetic calorimeter (CEM) module measures the 
transverse profile of the electromagnetic shower at a depth of approximately 
6 radiation lengths.  In front of each CEM module, a similar multiwire
chamber (the Central Preradiator, CPR) samples electromagnetic
showers that begin in the coil of the solenoid magnet.

The photon trigger used to acquire these data consists of three
levels \cite{trigger}.  At the first level, a single 
trigger tower~\cite{tower} in the CEM
is required to be above a threshold, typically $P_T>8$~GeV/c.  The
second trigger level finds clusters of transverse energy among the trigger
towers in the calorimeter, and requires that 88\% of the cluster
transverse energy be in the electromagnetic (EM) compartment of the
calorimeter.  In addition, the Level 2 electronics require that the transverse
energy in the 3~$\times$~3 grid of trigger towers surrounding the
photon candidate (equivalent to a radius {$R=\sqrt{(\Delta\eta)^2 +
(\Delta\phi)^2}=0.4$}) is less than 4~GeV, thereby requiring the
photon to be {\it isolated}~\cite{change_in_L2}.  
In the third level of the trigger, software algorithms require a CES cluster 
of energy of more than 0.5 GeV.  This cluster determines the 
position of the photon; fiducial cuts are then applied to avoid
uninstrumented regions of the detector.  
In both the second and third trigger levels, $P_T$ thresholds are applied.  
The data set at 1.8~TeV was selected by a  
prescaled \cite{prescale} trigger with a threshold of $P_T>10$~GeV/c,
an unprescaled trigger with a threshold of $P_T>23$~GeV/c,
and an unprescaled trigger
with a threshold of $P_T>50$~GeV/c without the isolation cut.  The
0.63~TeV data were acquired with an unprescaled trigger with a threshold of
$P_T>6$~GeV/c.  The respective integrated luminosities for the 1.8 TeV
data are 84,\ 84,\ and 1.1 pb$^{-1}$ for the  50,\ 23,\ and 10~GeV/c
thresholds, and 0.54 pb$^{-1}$ for the 0.63 TeV sample.

%
%

The selection of prompt photon candidates from the triggered events
is essentially the same as those used previously~\cite{pho93}, with
a minor change in the isolation cut~\cite{change_in_iso}.  The
selection cuts, cut efficiencies, and systematic
uncertainties are listed in Table~\ref{table:1b_eff}.   
Candidates are rejected if there is a
reconstructed charged track with $P_T$ greater 
than approximately 0.4 GeV/c pointing at the EM cluster 
or the CPR chamber containing the photon.  To improve the
signal/background ratio, the isolation cut applied in the trigger is
tightened to require less than 1~GeV of transverse energy in a cone
radius of 0.4.  After these selections, the main backgrounds to the prompt
photons are from single $\pi^0$ and $\eta$ mesons, with smaller
backgrounds from multiple $\pi^0$'s.  These backgrounds are
reduced by requiring there be no other photon candidate above 1~GeV
energy in the CES chamber containing the photon candidate.  The total
acceptance-times-efficiency for prompt photons within $|\eta|<0.9$ is
approximately 34\% for the 1.8 TeV data and approximately 37\% for the
0.63 TeV data (see Table 1).

%
%

\section{Background subtraction}
We employ two methods
for statistically subtracting the remaining neutral meson 
background from the photon candidates:
the {\em conversion method} counts the fraction of 
photon conversions ($\gamma \rightarrow e^+ e^-$) 
in the material of the magnet coil by using the CPR,
and the {\em profile method} uses the transverse 
profile of the electromagnetic shower in the CES.
For the conversion method, the probability of a single photon
conversion is approximately 60\%, while that for the 
two-photon decay of a $\pi^0$ or $\eta$ is larger, 
approximately 86\%. 
For the profile method, the transverse profile of the electromagnetic
shower of each photon 
candidate is compared to that measured
for electrons in a test beam in the same momentum range.  
On a statistical basis, a measure of the goodness of fit 
(labeled \zz since the
distribution for single photons is approximately a $\chi^2$ 
distribution)
is expected to be larger for a neutral meson decaying to two photons
than for a single photon because a neutral meson usually 
produces a wider EM shower~\cite{pho89}.  At 20 GeV/c 
approximately 80\% of the single photons and 40\% of the background
have a \zz \ less than 4. 
The conversion method has the advantage of an unlimited $P_T$ range
and smaller systematic uncertainties in the shape of the cross section
as a function of photon $P_T$.
The profile method has the advantage of a better 
separation of signal and background than the 
conversion method in the low $P_T$ region, where the two photons from
$\pi^0$ decay have a larger spatial separation.

%
%

For both background subtraction methods, 
the number of photons ($N_{\gamma}$) in a bin of $P_T$ is obtained 
from the number of photon candidates ($N$), the fraction of 
photon candidates that pass a fixed cut defined below
($\epsilon$), and the corresponding fractions for true
photons ($\epsilon_{\gamma}$) and background  ($\epsilon_b$), using:
\begin{equation}
\label{eq_method}
N_{\gamma} = \left(\frac{\epsilon - \epsilon_b}{\epsilon_{\gamma} - 
\epsilon_b}\right)N
\end{equation}

%
%
\noindent Equation~\ref{eq_method} comes from 
$\epsilon N = \epsilon_{\gamma} N_{\gamma} + \epsilon_b N_b$ with  
$N_b = N - N_{\gamma}$.
For the conversion method, $\epsilon$ is the fraction of photon candidates 
which produce a pulse height in the CPR greater than 1 minimum ionizing 
particle, within a 66 milliradian ``window'' (5 channels) 
centered on the photon direction in $\phi$.   
For the profile method, 
$\epsilon$ is the fraction of events which have 
$\tilde{\chi}^2<4$ out of all events with $\tilde{\chi}^2<20$.
Using these methods, we measure the signal/background 
ratio bin-by-bin and propagate the statistical uncertainty of each bin 
into 
the cross section measurement, including the effect of the
background subtraction.

%
%
The signal and background efficiencies for the two methods are 
similar to the previous analyses ~\cite{pho89,pho93}.   For the profile method 
$\es$ and $\eb$ are the same as those used in  
reference~\cite{pho89}.  For the conversion method $\epsilon_\gamma$ 
is estimated from the following equation:

$\es$~=~1~-~exp(-7t/9)

\noindent where t is the amount of material in radiation lengths 
in front of the CPR.  Corrections to this estimate of $\es$ are made
on an event-by-event basis for the amount of material
traversed, as well as changes in the pair
production cross section with photon energy~\cite{tsai}.  An
additional correction is made for photon showers that begin after the
photon has passed through the CPR, but in which a soft photon or electron from
the shower is scattered backwards at a large angle and gives a CPR
signal.  This correction is estimated with an electromagnetic shower
simulation~\cite{geant}. The final correction,
estimated using non-isolated photon triggers, 
is due to CPR signals arising from
soft photons from the underlying event or additional $\bar p p$ 
interactions (pileup). 
 With the higher luminosity of the current data sample, 
the number of pileup events increased;  hence, this correction is the
main difference between the current and previous
analyses~\cite{pho93}.   All of the corrections to $\es$ are 
applied to $\eb$ as well.  In addition $\eb$ is corrected for the multiple 
photons from background:
\begin{center}
$\eb$~=~1~-~exp(-7/9~*~t~*~$N_\gamma(P_T)$). 
\end{center}
\noindent The function $N_\gamma(P_T)$ is 
the average number of photons within the CPR ``window'' defined earlier.
This changes with transverse momentum and particle type and is 
estimated using a 
detector simulation of $\pi^0$, $\eta$ and $K_{S}^0$ mesons with a
relative production ratio of 1:1:$0.4$~\cite{pho89}.

%
%
The two methods developed to check the measured number of photons 
in the previous analysis~\cite{pho93} are repeated 
in this analysis.   The most important of these is the comparison 
of the number of photons determined by the conversion and 
profile methods in the region of photon $P_T$ where both methods are valid.   
The two methods should agree within their independent systematic 
uncertainties, estimated in the previous analysis to be 20\% for the 
profile method and 10\% for the conversion method for photon $P_T$ of 
20 GeV/c.   The two methods agreed to within 2\% in the previous 
analysis~\cite{pho93}, much better than expected given the 
systematic uncertainties.  
In the current analysis, before recalibration, the conversion method 
measurement of the photon cross section is 20\% smaller 
than the profile method measurement at $P_T$=20 GeV/c.  

\begin{figure}[ptbh]
\begin{center}
\begin{minipage}[h]{4.0in}
\epsfig{file=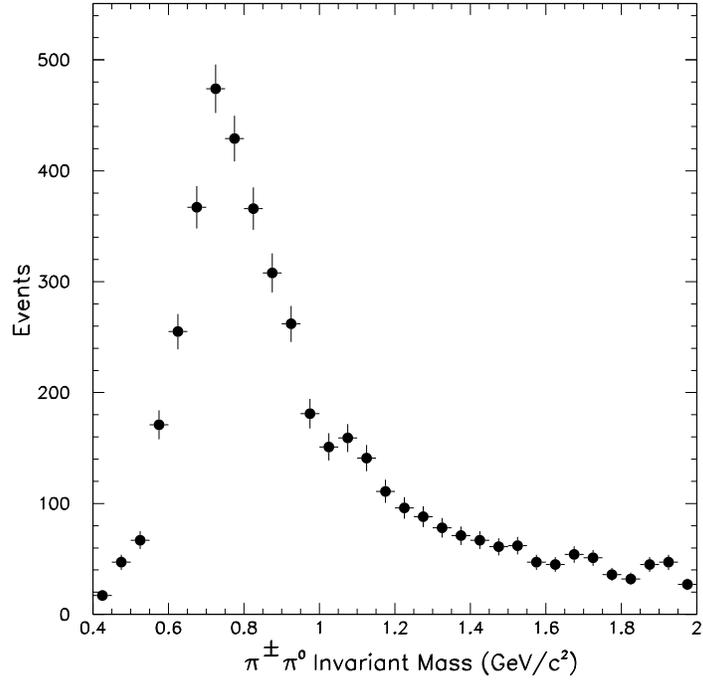,width=\linewidth}\caption{The 
$\rho^{\pm}$ data sample used to calibrate the profile and conversion 
methods.  Fits to the mass distribution using a Breit-Wigner line 
shape, plus a flat background component, failed to describe the data.
A gaussian fit to the truncated peak region gave a fitted mass of 0.767 GeV, 
consistent with the PDG $\rho$ meson mass.
As discussed in the appendix, a fit is not used in the analysis 
since the conversion method hit rate is the same for the 
signal and background dominated regions.}
\label{fig:rho}
\end{minipage}
\end{center}
\par
\end{figure}

The second method developed to check the two background subtraction 
methods uses reconstructed $\rho^\pm$ mesons (Figure~\ref{fig:rho}).   
Charged pions from the $\rho$ meson decays are required to 
fall in a separate calorimeter module from the neutral pions, which
then provide a clean probe of the photon measurement techniques, in 
particular $\eb$.  There was excellent agreement between the measured and 
expected efficiencies in the $\rho$ meson peak region in the previous analysis.  
There is also good agreement in the current profile method analysis,  
independently checked in the 1.8 TeV and 0.63 TeV data samples.
For example, in the 1.8 TeV sample the measured efficiency for 
the $\rho$ meson is $0.461 \pm 0.010$, and the expected value is 0.464.  
For the current conversion method analysis, however, there is 
a significant difference between the measured and expected hit rates, 
and the difference is identical in the 1.8 TeV and 0.63 TeV 
reconstructed $\rho$ meson data sets.  The measured conversion rate is 
$0.868 \pm .007$ ($0.814 \pm 0.008$), and the expected 
value is $0.835 \pm 0.006$ ($0.781 \pm 0.006$) in the 
1.8 (0.63) TeV $\rho$ meson data sample. 
After $\eb$ has been changed to agree with the measured hit rate from the  
$\rho$ meson sample, the 
the profile method and conversion method cross 
sections agree to within 10\% at $P_T$=20 GeV/c.  
The precise cause of this difference is not
understood, and its effect on the conversion method signal efficiency 
is unknown.  Therefore, we use the profile method cross section 
normalization at low $P_T$ to estimate the appropriate change 
in conversion method signal efficiency.  Hence,  
in this analysis the profile method determines the overall cross 
section normalization while the conversion method determines the 
cross section shape versus photon $P_T$.   This utilizes the 
strengths of each method 
for the combined cross section result. 
Further discussion 
of the photon background subtraction in this analysis can be 
found in the Appendix.   

\begin{figure}[ptbh]
\begin{center}
\begin{minipage}[h]{4.0in}
\epsfig{file=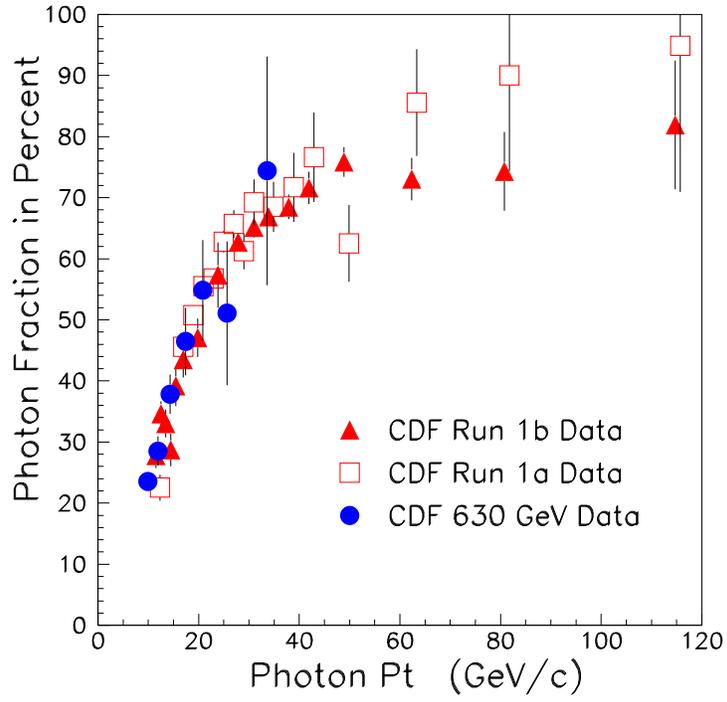,width=\linewidth}\caption{The 
photon fractions (ratio of the number of prompt photons to the 
number of prompt photon candidates) 
at the two different center-of-mass energies, and 
from the last published CDF analysis~\cite{pho93}.}%
\label{fig:phofrac}
\end{minipage}
\end{center}
\par
\end{figure}

Using the procedure outlined above, the purity
of the sample (number of photons/number of candidate events) is 
shown in Figure~\ref{fig:phofrac} as a function of photon $P_T$ for
the 1.8 TeV and 0.63 TeV data sets, as well as for the previous 1.8 TeV
analysis~\cite{pho93}.  The purity improves with increasing $P_T$ as expected 
from the enhanced effectiveness of the isolation cuts in reducing
jet backgrounds.  The differences in photon fraction at high $P_T$ 
between the two 1.8 TeV analyses may be due to changes in the 
number of background events due to differing isolation cuts.
Understanding such subtle effects in jet fragmentation is beyond the scope
of this paper, and the fraction of background events in a particular 
sample does not affect the prompt photon signal itself.

%
%
\section{Experimental systematic uncertainties}
The systematic uncertainties in the cross section normalization and shape
are dominated by uncertainties in $\es$ and $\eb$ for the background 
subtraction method for both cross sections at a given $P_T$.  The largest 
impact on the physics results presented
later in this article would be a systematic effect on the shape
of the cross section as a function of photon $P_T$.   Despite the difficulties 
with the conversion method normalization, the shape of the
measured cross section is well determined.  For example, the uncertainty 
due to the choice of $\es$ in the conversion method (as discussed 
in the Appendix) leads to a 12\% normalization uncertainty 
but only a 5\% change in slope between photon $P_T$ values of 
11 GeV/c to 115 GeV/c.  
The second uncertainty in $\es$ and $\eb$ comes from the correction 
for possible CPR hits from backscattered low energy photons and electrons
in the electromagnetic calorimeter shower. This is estimated 
with a detector simulation~\cite{geant} to be
1.6\% at 9~GeV/c and 6\% at 100~GeV/c.  
The uncertainty in the composition of the background~\cite{pho89} leads to a 
cross section uncertainty of 12\% at 9 GeV/c and 0.4\% at 100 GeV/c.
The entire mix of background sources 
is checked by a sample of 
events passing the same photon cuts as the data, with the 
exception of a slightly relaxed isolation
cut.  This shows agreement with expectations within
the uncertainty on $\eb$ quoted above.  
Finally, there are additional (correlated) uncertainties due to 
luminosity (4.1\% at 1.8~TeV and 4.4\% at 0.63~TeV),  
trigger efficiencies (2.2\% at 9 GeV/c and 5.4\% at 50 GeV/c), 
selection efficiencies (3.6\% at 1.8~TeV and 6.2\% at 0.63~TeV), 
and photon energy scale (4.5\%).

%
%
\section{Cross section evaluation and comparison to QCD models}
From the number of prompt photons 
in a given bin of transverse
momentum, along with the  
acceptance and the integrated luminosity for that bin, 
the isolated prompt photon cross section is derived
and tabulated in Tables~2 and 3.  
Also tabulated are the
number of events (photon candidates), the number of photons after 
the background subtraction, and the 
statistical and systematic uncertainties.  The systematic uncertainties
listed are approximately 100\% correlated bin-to-bin and 
include all normalization uncertainties.  

\begin{figure}[ptbh]
\begin{center}
\begin{minipage}[h]{4.0in}
\epsfig{file=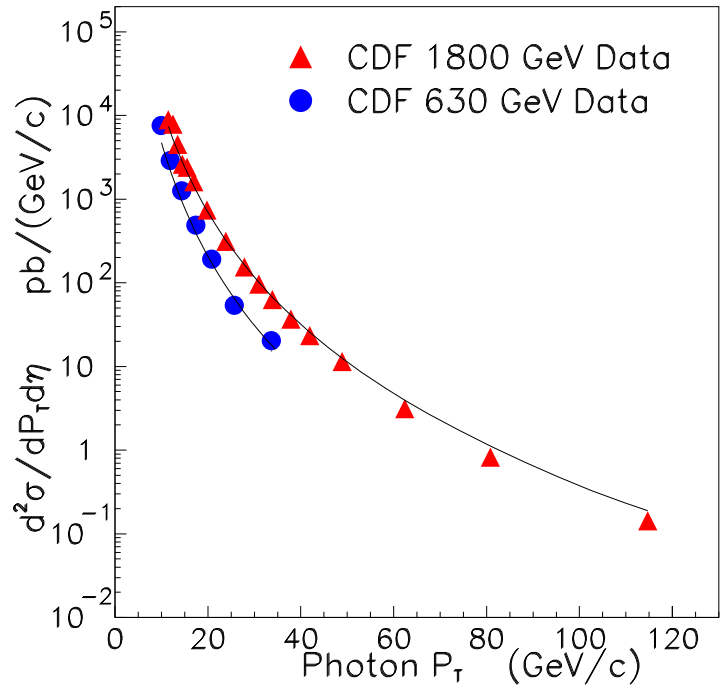,width=\linewidth}\caption{The 
inclusive photon cross sections at the two different 
center-of-mass energies compared to the
next-to-leading order QCD predictions of ref.~\cite{nloqcd}. }%
\label{fig:log}
\end{minipage}
\end{center}
\par
\end{figure}

\begin{figure}[ptbh]
\epsfig{file=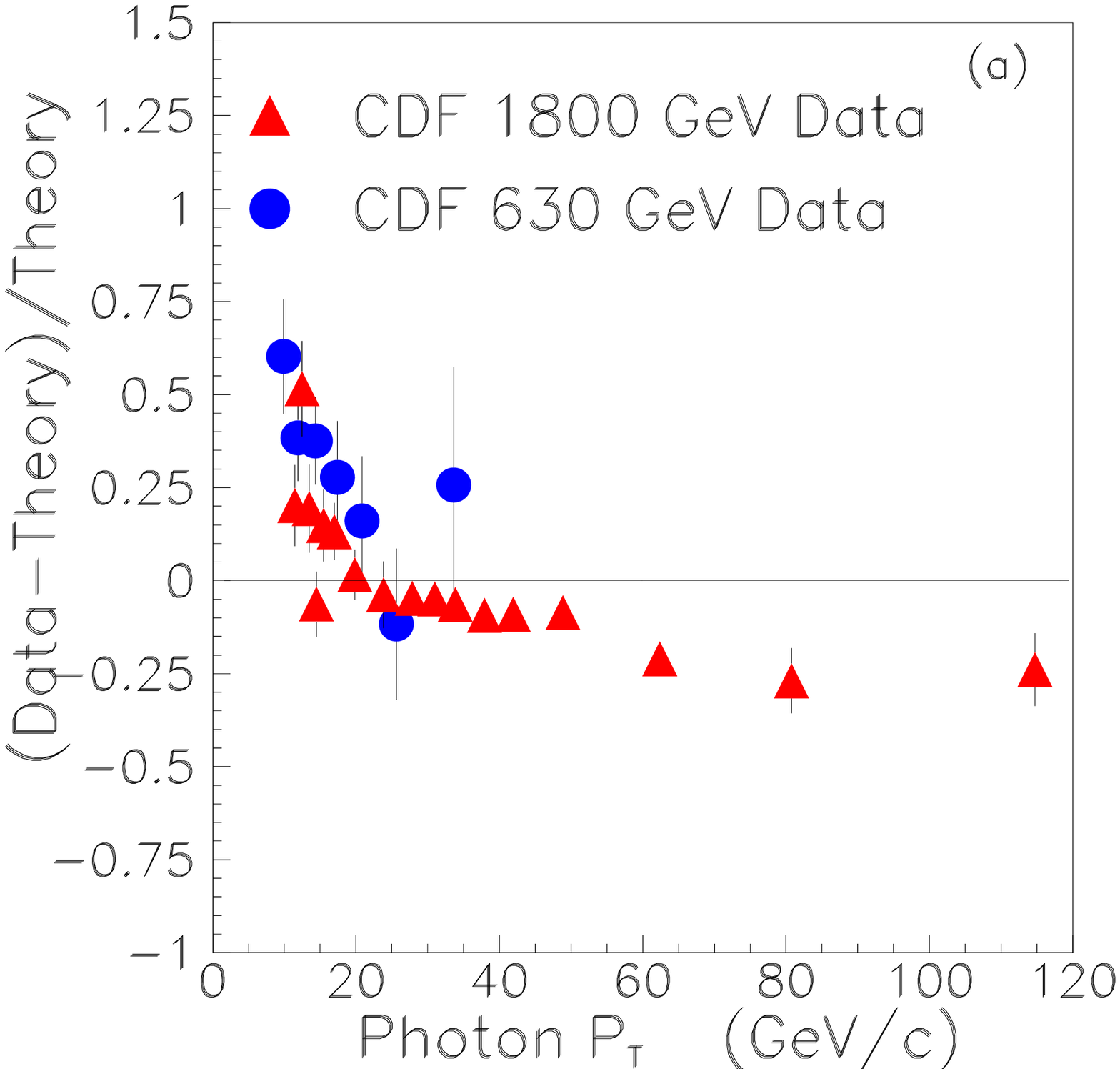,width=0.5\linewidth}\label{fig:ptpt} \epsfig
{file=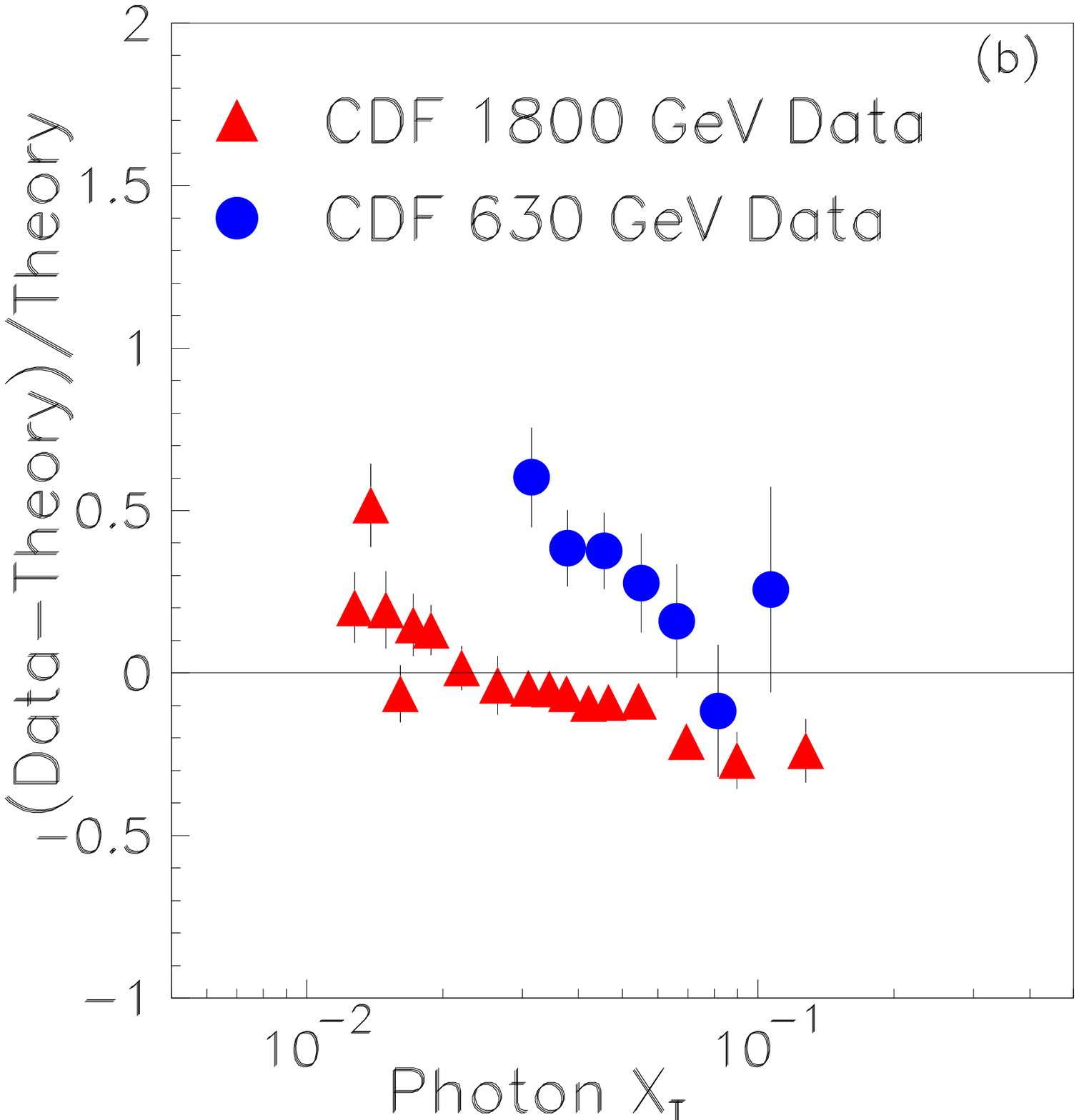,width=0.5\linewidth}\caption{A comparison of the 1.8 TeV
 and 0.63 TeV data to a NLO QCD calculation~\cite{nloqcd} as a function of 
photon $P_T$ and $x_T$. 
The NLO QCD calculation used the CTEQ5M parton distributions and a 
scale of $\mu=P_T$.
}\label{fig:ptxt}\end{figure}

\begin{figure}[ptbh]
\epsfig{file=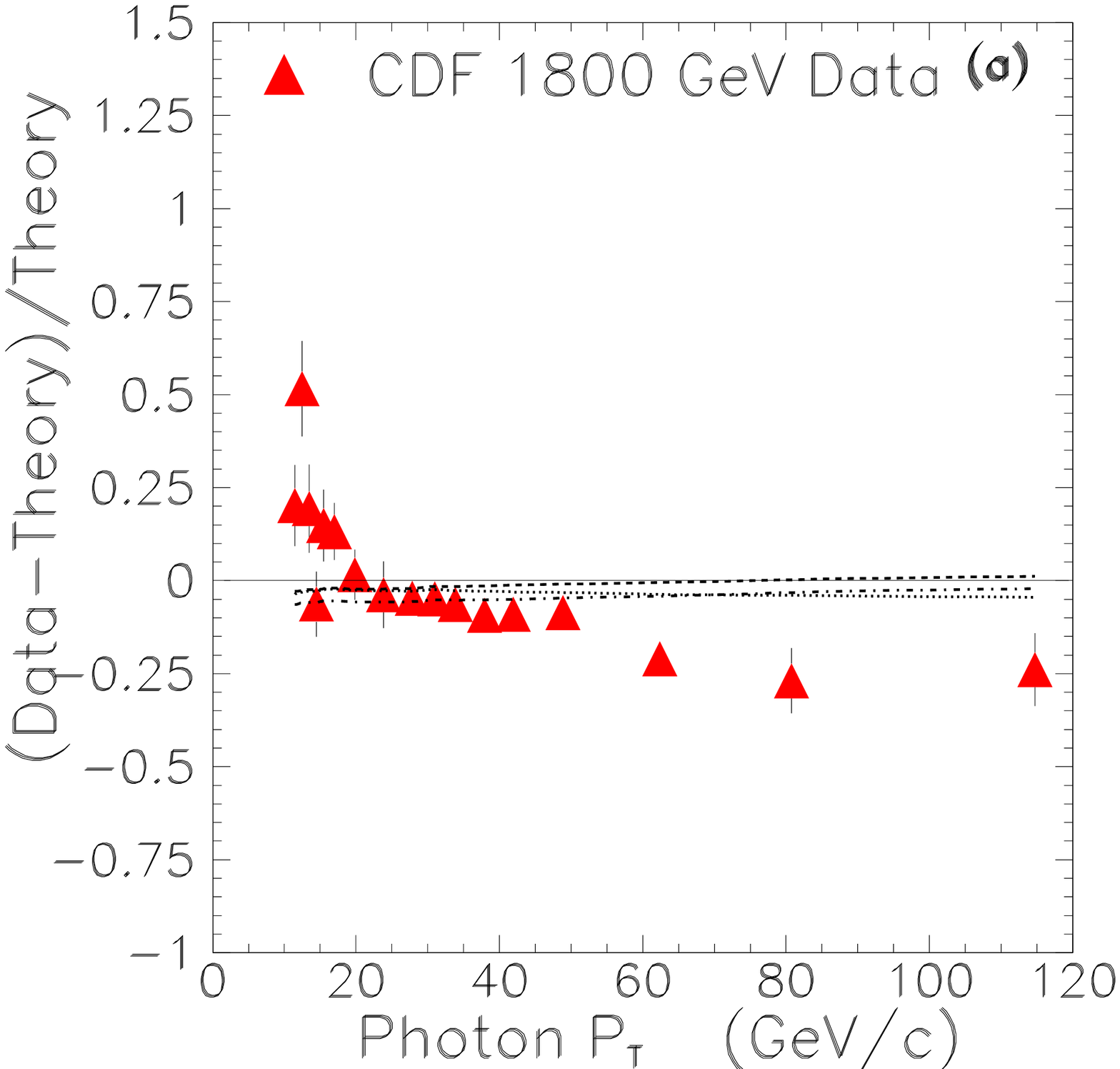,width=0.5\linewidth} \epsfig
{file=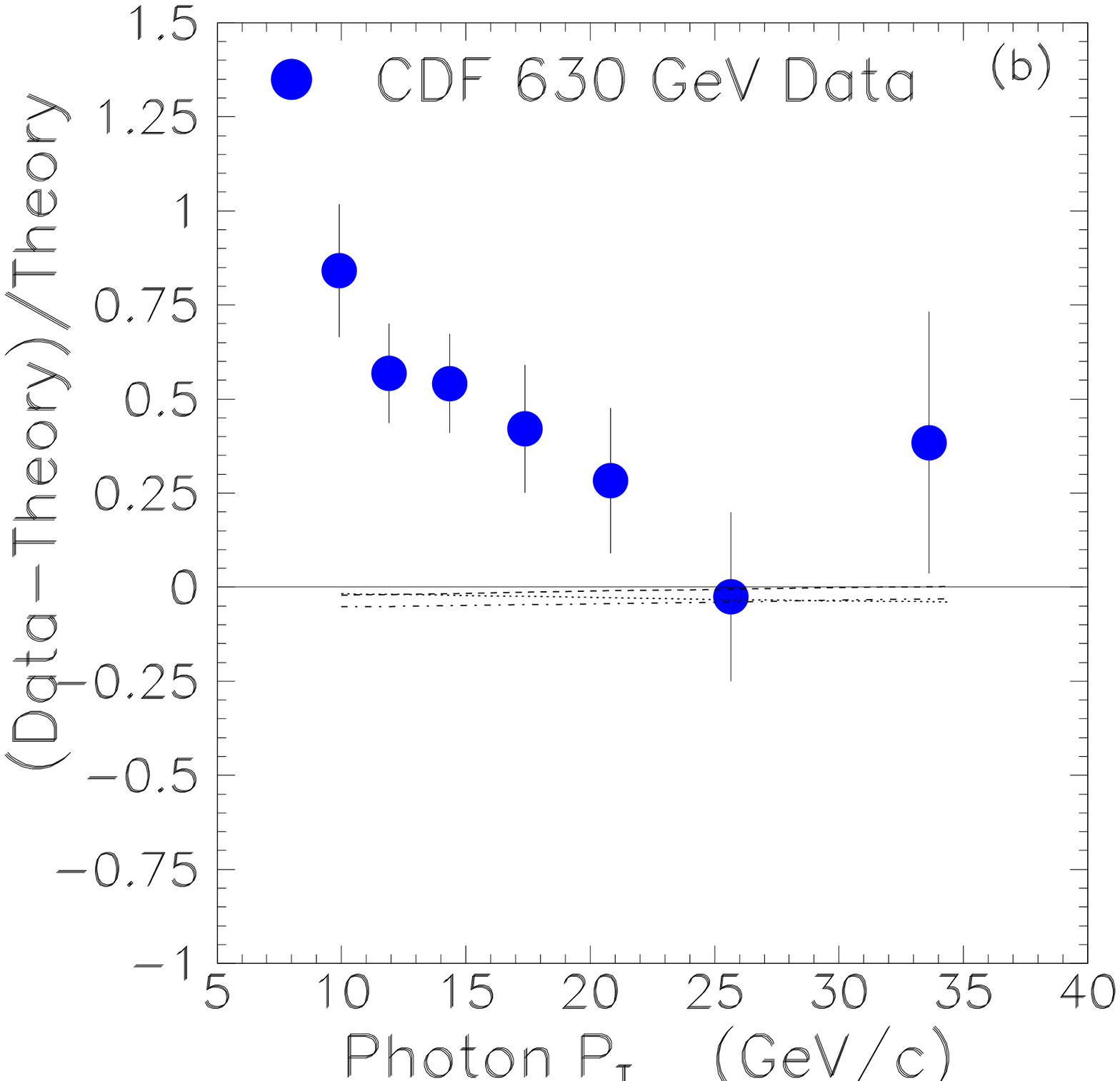,width=0.5\linewidth}\caption{A comparison 
of the 1.8 TeV and 0.63 TeV differential cross sections
 to NLO QCD calculations 
using different parton distribution functions: 
CTEQ5M~\cite{cteq5} (solid line), CTEQ5HJ~\cite{cteq5} (dashed line), 
MRST-99~\cite{mrst} (dotted line), 
MRST-99 $g\uparrow$~\cite{mrst} (dot-dash line). 
All calculations use a scale of $\mu=P_T$. 
}\label{fig:pdf_comp}\end{figure}

\begin{figure}[ptbh]
\epsfig{file=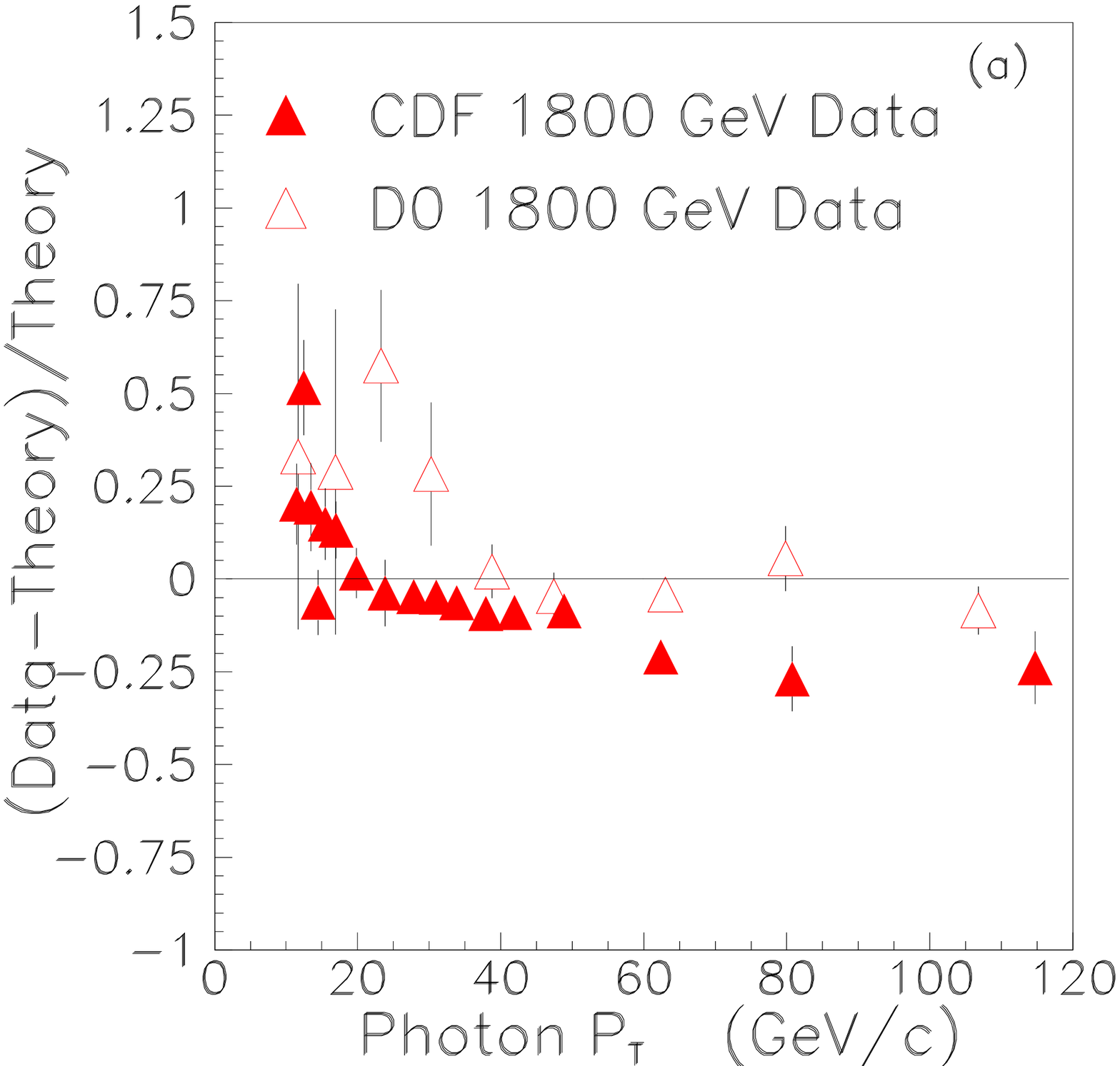,width=0.5\linewidth}\label{fig:d0pt} \epsfig
{file=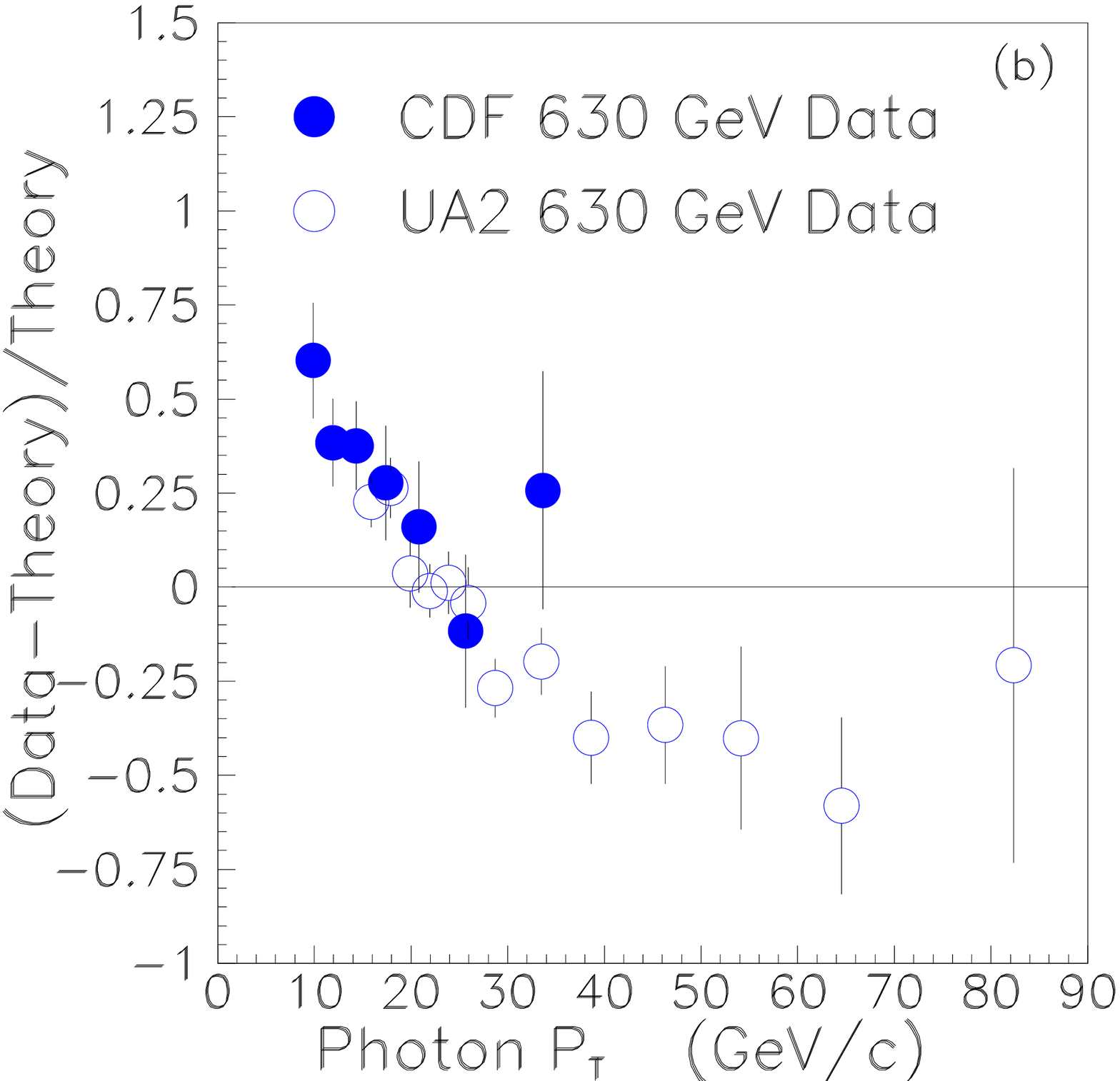,width=0.5\linewidth}\caption{A comparison of the 
CDF and D0 1.8 TeV data sets and the CDF and UA2 630 GeV data sets to 
the same NLO QCD calculation~\cite{nloqcd} as Figure~\ref{fig:ptxt}.
There is excellent agreement between the CDF and UA2 data sets.  The 
CDF and D0 data sets differ in normalization by about 20\%,  consistent 
with the quoted correlated systematic uncertainties of the measurements.
The correlated systematic uncertainties for the D0 data 
are 10\% at large $P_T$ growing
to 74\% in the lowest $P_T$ bin.  CDF's correlated systematic uncertainties 
are listed in Tables 2 and 3, and are 11\% at large $P_T$ growing to 
18\% in the lowest $P_T$ bin.
 }\label{fig:ua2pt}\end{figure}

%
%
In Figure~\ref{fig:log} measurements from both data sets are compared to 
a next-to-leading order QCD calculation~\cite{nloqcd}
derived using the CTEQ5M parton 
distributions~\cite{cteq5} and $\mu=P_T$ for the renormalization, 
factorization, and fragmentation scales.  
The QCD prediction agrees qualitatively with the measurements over
more than 4 orders of magnitude in cross section.  
Figure~\ref{fig:ptxt}a shows a
comparison of the cross sections
as a function of photon $P_T$ and Figure~\ref{fig:ptxt}b shows a 
comparison of the cross sections as a 
function of the scaling variable $x_T (\equiv 2P_T/\sqrt{s}$).
The shape of the cross sections versus $P_T$ (or $x_T$) is generally 
steeper than that of the theoretical predictions.  
Many variations of modern parton distributions and 
scales were tried, with small changes in the shape of the 
predictions, but none accurately predicted the 
shape of the cross sections, as seen in Figure~\ref{fig:pdf_comp}.  
It is still possible other changes in the theory parameters 
could improve the agreement between data and theory for one or
the other data sets.
However, the comparison of the two cross sections   
as a function of photon $x_T$, a ratio in 
which most experimental and theoretical uncertainties
cancel,  is more difficult to reconcile with the NLO QCD calculations.
The parton distributions at a fixed value of $x_T$ are 
the same for 1.8 TeV and 0.63 TeV center-of-mass energies, 
except for changes due to QCD evolution which are very
similar for different parton distribution parameterizations.
In the region where the 1.8 TeV and
0.63 TeV data sets overlap in $x_T$,  the variation of the 
0.63 TeV/1.8 TeV cross section ratio
with parton distribution~\cite{pdf} is 1\%,  and the variation with 
scale (between $\mu$=$P_T$ and $\mu$=$P_T/2$) is only 4\%.  
The experimental systematic uncertainties are also smaller in the 
$x_T$ ratio,  with the quadrature sum of the uncertainties
in the 0.63 TeV/1.8 TeV ratio reduced to 10\% at $x_T$=0.03
and  5\% at $x_T$=0.15.  
The measured ratio of 
cross sections, however, is more than 50\% larger than that 
predicted by NLO QCD (the ratio of data points shown 
in Figure~\ref{fig:ptxt}b), 
and the disagreement is essentially independent of $x_T$ in the range 
where 1.8 TeV and 0.63 TeV data sets
overlap.  The ratio is more
than four standard deviations larger than that predicted
by current NLO QCD calculations. 
These results are reinforced when the CDF cross sections are 
compared to the results of the D0 and UA2 experiments~\cite{d0,ua2} as 
shown in Figure~\ref{fig:ua2pt}.  
There is excellent agreement between the CDF and UA2 data sets.  The 
CDF and D0 data sets differ in normalization by about 20\%,  consistent 
with the quoted correlated systematic uncertainties of the measurements.
The correlated systematic uncertainties of the D0 measurement 
are 10\% at large $P_T$ growing
to 74\% in the lowest $P_T$ bin.  CDF's correlated systematic uncertainties 
are listed in Tables 2 and 3, and are 11\% at large $P_T$ growing to 
18\% in the lowest $P_T$ bin.  Recently the D0 experiment has 
published a measurement of the photon cross section at 0.63 TeV, 
as well as the ratio of 0.63 TeV and 1.8 TeV cross sections~\cite{d0rat}.
In the D0 ratio measurement the lowest $x_T$ points are systematically
higher than NLO QCD, but the deviations are not significant
in light of the combined statistical and systematic uncertainties.

One possibility for the observed discrepancy with NLO QCD is the 
lack of a complete description of the initial state parton 
shower in the NLO QCD calculation, which
could give a recoil effect to the photon+jet system (``$k_T$'').  
Higher order QCD calculations including such effects are 
becoming available~\cite{resum}, but are not ready for detailed 
comparisons at this time.  
To explore qualitatively the effect of $k_T$ on the comparisons, 
we have added a simplified gaussian smearing 
to the NLO QCD calculations to see if the 
measurements could be sensitive to these effects.  The photon+jet 
system was given a transverse momentum recoil consistent with that
measured in the Drell-Yan process at each center-of-mass energy (3 GeV
at 0.63 TeV and 4 GeV at 1.8 TeV).   The comparisons with 
the measurements are improved with the addition of these amounts of 
$k_T$.  For example, the measured ratio of cross sections versus
$x_T$ is only 19\% larger than NLO QCD + $k_T$, compared to 
the 50\% excess without $k_T$ discussed earlier.  We look forward
to the maturation of the QCD calculations including the recoil 
effect due to soft gluon radiation.

\section{Acknowledgments}
We would like to thank Werner Vogelsang for 
providing the theoretical calculations used in this paper.
We also thank the Fermilab staff and the technical staffs of the
participating institutions for their vital contributions.  This work was
supported by the U.S. Department of Energy and National Science Foundation;
the Italian Istituto Nazionale di Fisica Nucleare; the Ministry of Education,
Science, Sports and Culture of Japan; the Natural Sciences and Engineering 
Research Council of Canada; the National Science Council of the Republic of 
China; the Swiss National Science Foundation; the A. P. Sloan Foundation; the
Bundesministerium fuer Bildung und Forschung, Germany; the Korea Science 
and Engineering Foundation (KoSEF), the Korea Research Foundation, and the 
Comision Interministerial de Ciencia y Tecnologia, Spain.

\newpage

\begin{center}
\large{\bf{Appendix: Recalibration of the Photon Background Subtraction}}
\end{center}

\begin{figure}[ptbh]
\epsfig{file=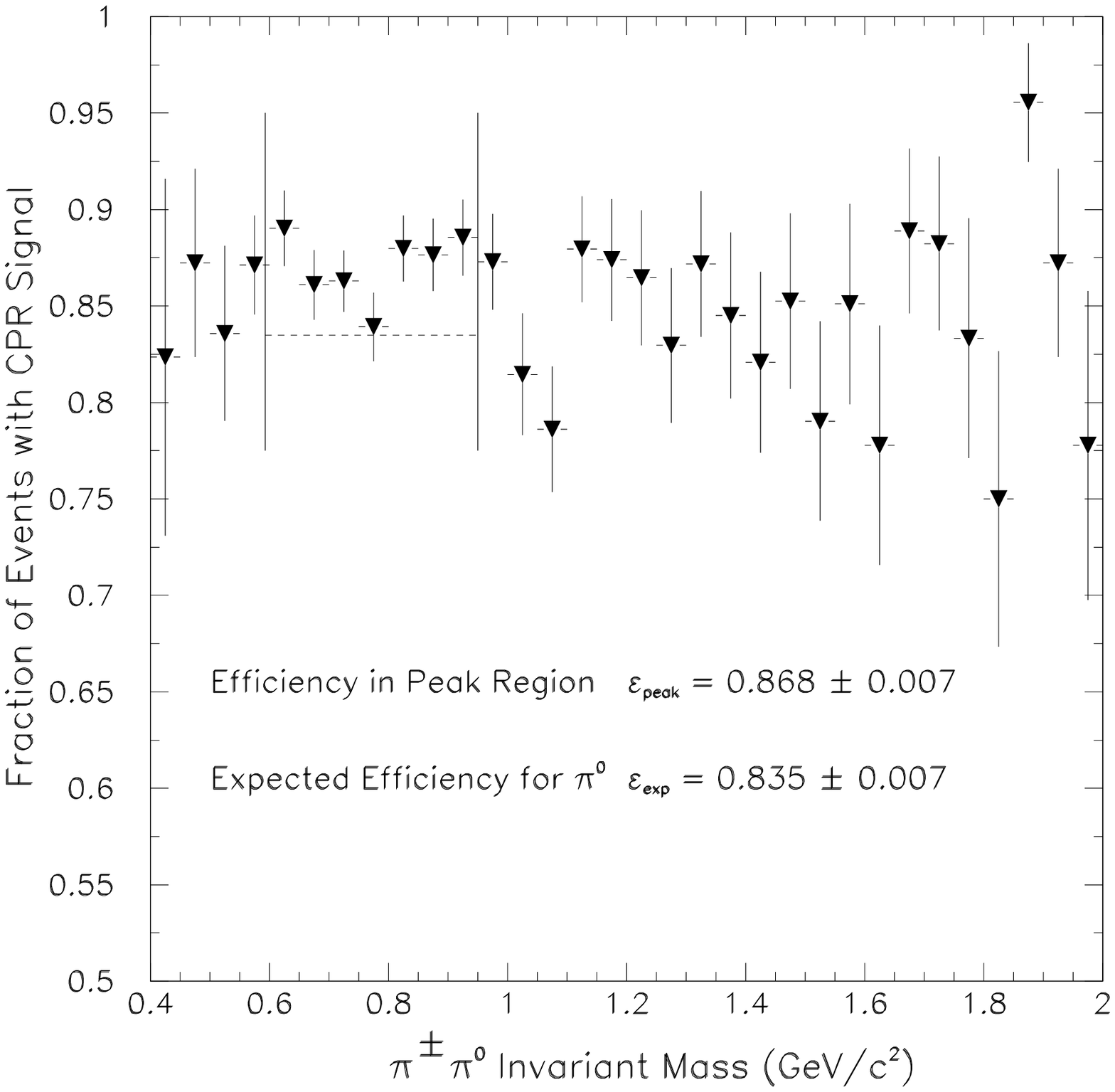,width=\linewidth}
\caption{The fraction of events with a CPR signal is shown as 
a function of measured invariant mass in the $\rho^{\pm}$ sample.
The dashed line shows the expected $\pi^0$ CPR signal rate in the
$\rho$ meson peak region.
 }\label{fig:effmass}\end{figure}

As mentioned in the main text of this paper, a sample of reconstructed 
$\pi^0$s from charged $\rho$ meson decays has been used to check 
the two techniques for subtracting photon backgrounds.  The 
relatively pure measurements of $\pi^0$s agree well with expectations
for the profile method, but do not for the conversion method,
as shown in figure~\ref{fig:effmass}.  The dashed line in this
figure is the expected $\pi^0$ CPR signal rate in the $\rho$ meson peak 
region,  falling below the data.  The signal rate is largely independent
of measured mass, and sideband subtractions of the hit rate for
various selected mass regions had no effect on this result.
The three most likely sources for the conversion method discrepancy,
which have been extensively investigated~\cite{dsp}, are:
1) an underestimate of the 
correction for CPR hits from multiple $p\bar p$ collisions; 
2) an underestimate of the material in front of the CPR chambers; 
3) a change in CPR chamber performance compared to the 1992 analysis.    
Our analysis has shown that no single source 
is the likely cause of the discrepancy; it is perhaps a complicated 
mixture of multiple effects~\cite{dsp}.  The $\rho$ meson 
sample is therefore used to 
recalibrate the conversion method.
When this recalibration is done by correcting for
the difference in the measured and expected $\rho$ meson conversion rates, 
agreement is restored between 
the conversion method and profile method photon cross section measurements 
at both 1.8 TeV and 0.63 TeV.  
To do this, however, one has to assume  
a correction for the signal efficiency as well as the background efficiency, 
since the two are usually correlated.  
The size of the signal efficiency correction 
depends on the source of the conversion 
method discrepancy.  The most extreme choice for the change in signal 
efficiency,
which comes from the assumption that the entire source of the 
discrepancy is multiple $p\bar p$ collisions, 
increases the conversion method normalization by 30\% more than no
correction at all.  On the other hand, 
the profile method normalization is apparently very robust at low $P_T$. 
As an example,  when the efficiencies in the profile method are changed based
on the measured uncertainty in the $\rho$ meson
 measurement,  the cross section 
changes by only 5\% at $P_T=$20 GeV/c.  Therefore, in the final cross section 
measurement the two methods are 
combined based on their respective strengths.  The profile method 
determines the normalization of the cross section at low $P_T$, while 
the conversion method determines the shape with photon $P_T$.  This is
accomplished by correcting the conversion method background efficiency 
directly with the $\rho$ meson measurement,  then choosing the signal 
efficiency that matches the profile method normalization at low $P_T$.

\bigskip%
\begin{table}[pth]
\begin{center}
\begin{tabular}{|l|c|c|c|c|}\hline 
 & \multicolumn{2}{|c|}{1.8~TeV} &  \multicolumn{2}{|c|}{0.63~TeV}\\
Analysis Cut & Efficiency   & Uncertainty & Efficiency   & Uncertainty\\ \hline
Remove Uninstrumented Regions & 0.977 & 0.010  & 0.977 & 0.010 \\ 
Fiducial      & 0.64  & 0.000  & 0.64  & 0.000 \\
$|Z_{vertex}| <$ 60 cm   & 0.937 & 0.011 & 0.85 & 0.05 \\
Isolation    & 0.832 & 0.004 & 0.919 & 0.013\\ 
No track   & 0.797 & 0.005 & 0.853 & 0.007 \\
Energy in $2^{nd}$ CES cluster $<$ 1 GeV  & 0.893 & 0.031 & 0.893 & 0.031\\
Missing $E_{T}$/$E_T <$0.8 & 0.976 & 0.014 & 0.976 & 0.014 \\ \hline
Total & 0.339 & 0.036 & 0.372 & 0.062\\ \hline
\end{tabular}
\caption{A list of the photon selection efficiencies and their 
uncertainties for the 1.8 and 0.63 TeV data.  The selection criteria and 
techniques used to measure their efficiencies are very similar to
the previous analysis~\cite{pho93}.  Several CES and 
CEM channels were not working and were removed.  The `fiducial' 
cuts require that the photon is not  
close to detector boundaries.  The `$Z_{vertex}$' cut requires the 
$\bar p p$ interaction point to be near the center of the detector. 
The `isolation' cut 
requires that the transverse energy be
less than 1 GeV in a cone of 
radius {$R=\sqrt{(\Delta\eta)^2 + (\Delta\phi)^2}=0.4$} around 
the photon direction.  The `no track' cut refers to events with a 
reconstructed track pointing at the CPR chamber containing the photon.  
The `energy in a 2nd CES cluster' cut reduces backgrounds from neutral 
mesons. The `missing $E_T$' cut removes photon candidates arising
from cosmic rays.}
\label{table:1b_eff}
\end{center}
\end{table}

%
%
\renewcommand{\baselinestretch}{1.6}
\begin{table}
\begin{center}
\begin{tabular}
[c]{|c|c|c|c|c|c|c|}\hline
\multicolumn{7}{|c|}{$\sqrt{s}=1.8$~TeV}\\\hline
${\small P}_{T}$ & {\small \# Candidates} & {\small \# Photons} &
${\small d}^{2}{\small \sigma/dP}_{T}{\small d\eta}$ &
{\small Stat.} & {\small Sys.} & {\small NLO QCD}\\\hline
{\small (GeV/c)} & N & $N_{\gamma}$ & {\small (pb/(GeV/c))} & {\small (\%)}%
& {\small (\%)} & {\small (pb/(GeV/c))}\\\hline
{\small 11.5} & 13818 & 3839 & {\small 8.84 }$\times${\small  10}$^{3}$%
& 9.0 & 18.0  & {\small 7.36 }$\times${\small  10}$^{3}$\\\hline
{\small 12.5} & 12809 & 4437 & {\small 7.89 }$\times${\small  10}$^{3}$%
& 8.5 & 14.4 & {\small 5.21 }$\times${\small  10}$^{3}$\\\hline
{\small 13.5} & 9304 & 3074 & {\small 4.50 }$\times${\small  10}$^{3}$%
& 10.0 & 14.5 & {\small 3.77}$\times${\small  10}$^{3}$\\\hline
{\small 14.5} & 6173 & 1772 & {\small 2.61 }$\times${\small  10}$^{3}$%
& 9.3 & 16.3 & {\small 2.78 }$\times${\small  10}$^{3}$\\\hline
{\small 15.5} & 4150 & 1626 & {\small 2.40 }$\times${\small  10}$^{3}$%
& 8.4 & 12.1 &{\small 2.09 }$\times${\small  10}$^{3}$ \\\hline
{\small 17.0} & 4993 & 2173 & {\small 1.61 }$\times${\small  10}$^{3}$%
& 6.8 & 12.5 & {\small 1.43 }$\times${\small  10}$^{3}$\\\hline
{\small 19.8} & 4133 & 1945 & {\small 7.38 }$\times${\small  10}$^{2}$%
& 6.7 & 12.0 & {\small 7.26 }$\times${\small  10}$^{2}$\\\hline
{\small 23.8} & 1410 & 809 & {\small 3.12 }$\times${\small  10}$^{2}$%
& 9.3 & 11.3 &{\small 3.24 }$\times${\small  10}$^{2}$ \\\hline
{\small 27.9} & 38033 & 25226 & {\small 1.55 }$\times${\small  10}$^{2}$%
& 3.5 & 10.5 & {\small 1.63 }$\times${\small  10}$^{2}$\\\hline
{\small 31.0} & 13283 & 9171 & {\small 9.64 }$\times${\small  10}$^{1}$%
& 2.6 & 10.5 & {\small 1.01 }$\times${\small  10}$^{2}$\\\hline
{\small 33.9} & 16767 & 11885 & {\small 6.32 }$\times${\small  10}$^{1}$%
& 2.3 & 10.8 & {\small 6.75 }$\times${\small  10}$^{1}$\\\hline
{\small 37.9} & 9244 & 6750 & {\small 3.69 }$\times${\small  10}$^{1}$%
& 3.0 & 10.8 & {\small 4.06 }$\times${\small  10}$^{1}$\\\hline
{\small 41.9} & 5467 & 4210 & {\small 2.33 }$\times${\small  10}$^{1}$%
& 3.7 & 10.8 &{\small 2.57 }$\times${\small  10}$^{1}$ \\\hline
{\small 48.9} & 6683 & 5453 & {\small 1.14 }$\times${\small  10}$^{1}$%
& 3.3 &  11.3 & {\small 1.25 }$\times${\small  10}$^{1}$\\\hline
{\small 62.4} & 3253 & 2376 & {\small 3.12 }$\times${\small  10}$^{0}$%
& 4.8 & 10.2 & {\small 3.96 }$\times${\small  10}$^{0}$\\\hline
{\small 80.8} & 924 & 686 & {\small 8.21 }$\times${\small  10}$^{-1}$%
& 12.0 & 10.6 & {\small 1.12 }$\times${\small  10}$^{0}$\\\hline
{\small 114.7} & 386 & 316 & {\small 1.43 }$\times${\small  10}
$^{-1}$ & 13.0 & 11.4 & {\small 1.89 }$\times${\small  10}$^{-1}$\\\hline
\end{tabular}
\end{center}

\caption{The 1.8 TeV isolated photon cross section is tabulated 
along with the statistical and systematic uncertainties.
The systematic uncertainties include normalization uncertainties and are 
approximately 100\% correlated bin-to-bin.  The column labeled NLO QCD is the
result of the calculation discussed in ~\cite{nloqcd}.}
\end{table}
\bigskip

%
%
\renewcommand{\baselinestretch}{2.0}
\bigskip%
\begin{table}
\begin{center}
\begin{tabular}
[c]{|c|c|c|c|c|c|c|}\hline
\multicolumn{7}{|c|}{$\sqrt{s}=0.63$~TeV}\\\hline
${\small P}_{T}$ & {\small \# Candidates} & {\small \# Photons} &
${\small d}^{2}{\small \sigma/dP}_{T}{\small d\eta}$ &
{\small Stat.} & {\small Sys.} & {\small NLO QCD}\\\hline
{\small (GeV/c)} & N & $N_{\gamma}$ & {\small (pb/(GeV/c))} & {\small (\%)}%
& {\small (\%)} & {\small (pb/(GeV/c))}\\\hline
{\small 9.9} & 26606 & 6260 & {\small 7.55 }$\times${\small  10}$^{3}$ 
& 9.5 & 21.6 & {\small 4.71 }$\times${\small  10}$^{3}$\\\hline
{\small 11.9} & 8531 & 2382 & {\small 2.90 }$\times${\small  10}$^{3}$%
& 8.4 & 18.8 & {\small 2.09 }$\times${\small  10}$^{3}$\\\hline
{\small 14.3} & 4048 & 1532 & {\small 1.26 }$\times${\small  10}$^{3}$%
& 8.5 & 16.5 & {\small 9.12 }$\times${\small  10}$^{2}$\\\hline
{\small 17.4} & 1269 & 590 & {\small 4.89 }$\times${\small  10}$^{2}$%
& 11.9 & 15.6 & {\small 3.83 }$\times${\small  10}$^{2}$\\\hline
{\small 20.8} & 550 & 302 & {\small 1.92}$\times${\small  10}$^{2}$%
& 15.0 & 15.0 & {\small 1.66 }$\times${\small  10}$^{2}$\\\hline
{\small 25.7} & 245 & 125 & {\small 5.40 }$\times${\small  10}$^{1}$%
& 23.0 & 14.9 & {\small 6.11 }$\times${\small  10}$^{1}$\\\hline
{\small 33.6} & 112 & 84 & {\small 2.03 }$\times${\small  10}$^{1}$%
& 25.2 & 14.8 & {\small 1.61 }$\times${\small  10}$^{1}$\\\hline
\end{tabular}
\end{center}
\caption{The 0.63 TeV isolated photon cross section calculated is tabulated along with the statistical and systematic uncertainties.
The systematic uncertainties include normalization uncertainties and are 
approximately 100\% correlated bin-to-bin.   The column labeled NLO QCD is the
result of the calculation discussed in ~\cite{nloqcd}.}
\end{table}

%
%
\bigskip

%
%

%
%

\begin{thebibliography}{9}
\bibitem{change_in_L2} This is a change from the Level 2 configuration
used in Ref~\cite{pho93}, in which a 5$\times$5 array was used to
isolate the photon.
\bibitem{prescale} A prescaled trigger is one in which every Nth event
is recorded, where N is an integer, called the prescale factor.
\bibitem{change_in_iso} The isolation cut used in this analysis
measures energy in a cone of radius 0.4 around the photon candidate;
a cone radius of 0.7 was used in Ref~\cite{pho93}.
\bibitem{pdf} The other parton distribution functions used for comparison were CTEQ5HJ~\cite{cteq5}, MRST-99~\cite{mrst}, and MRST-99 $g\uparrow$~\cite{mrst}.
\end{thebibliography}
\end{document}